\documentstyle[aps,amssymb,epsf,twocolumn]{revtex}
\newcommand{\eins}{\mbox{$1
\hspace{-1.0mm} {\bf l}$}}
 \newcommand{\be}{\begin{equation}}
\newcommand{\ee}{\end{equation}}
\newcommand{\bea}{\begin{eqnarray}}
\newcommand{\eea}{\end{eqnarray}}
 \newcommand{\ket}[1]{ | \, #1
\rangle}
 \newcommand{\bra}[1]{ \langle #1 \, |}
\newcommand{\proj}[1]{\ket{#1}\bra{#1}}

\newcommand{\cala}{\mbox{$\cal A$}}
 \newcommand{\calb}{\mbox{$\cal
B$}}
\begin{document}
\title{Approximate quantum cloning \\ and the impossibility of 
superluminal information transfer}
\author{D. Bru\ss $^1$, G.M. D'Ariano$^2$, C. Macchiavello$^2$, and
M.F. Sacchi$^2$}
\address{$^1$Inst. f\"{u}r Theoret. Physik,
                 Universit\"{a}t Hannover, Appelstr. 2, D-30167
                 Hannover, Germany\\ $^2$Theoretical Quantum Optics
    Group, Dipartimento di Fisica
                 ``A. Volta'' \\ and INFM-Unit\`a di Pavia,
                 Via Bassi 6, 27100 Pavia, Italy}
\maketitle
\begin{abstract}
We show that non-locality of quantum mechanics cannot lead to
superluminal transmission of information, 
even if  most general local operations are allowed, as long as
they are linear and trace
preserving. In particular, any quantum
mechanical approximate cloning transformation does not allow
signalling.  
On the other hand, the no-signalling constraint on its own is
not sufficient to prevent a transformation from surpassing
the known cloning bounds.
We illustrate these concepts on the basis of some
examples.
\end{abstract}
\section{Introduction}
The impossibility of superluminal communication through the use of
quantum entanglement has already been vividly discussed in the past,
see for example \cite{grw,Herbert,Wootters,ghi,BuschA,svetlichny,Peres2}.  
Recently this topic has re-entered the stage of present research in
the context of quantum cloning: the no-signalling constraint has been
used to derive upper bounds for the fidelity of cloning
transformations \cite{Gisin,Hardy,ghosh,Pati}. 
As the connection between approximate cloning and no-signalling
is still widely debated, we aim at clarifying  in this paper the
quantum mechanical principles that forbid superluminal communication,
and at answering the question whether they are the same principles
that set limits to quantum cloning.

Our scenario throughout the paper for the attempt to transmit
information with superluminal speed is the well-known
entanglement-based communication 
scheme\cite{Herbert,Wootters,ghi}. The idea is 
the following: two space-like separated parties, say Alice and Bob,
share an entangled state of a pair of two-dimensional quantum systems
(qubits), for example the singlet state
$\ket{\psi_s}=(\ket{01}-\ket{10})/\sqrt{2}$. Alice encodes a bit of
information
 by choosing between two possible orthogonal measurement
bases for her qubit and performing
 the corresponding measurement. By
the reduction postulate, the qubit at Bob's side collapses into a pure
state depending on the result of the measurement performed by
Alice. If a perfect cloning machine were available, Bob could now
generate an infinite number of copies of his state, and therefore would
be able to determine his state with perfect accuracy, thus knowing
what basis Alice decided to use. In this way, transfer of information
between Alice and Bob would be possible. In particular, if they are
space-like separated, information could be transmitted with
superluminal speed.  The same transfer of information could evidently
also be
 obtained if it were possible to determine the state of a
single
 quantum system with perfect accuracy, which is also
impossible \cite{DarYu,Busch}.

One might ask the question whether approximate cloning allows
superluminal communication \cite{Ghirardi}: with imperfect cloning Bob
can produce a number of imperfect copies, and thus get some
information about his state. But this information is never enough to
learn Alice's direction of measurement. This has been shown in
Ref. \cite{zeil} for a specific example. More generally, as we will
show in this paper, the reason is that {\em no} local linear
transformation can lead to transmission of information through
entanglement, but any cloning operation consistent with quantum
mechanics has to be linear.

The fact that  non-locality of quantum entanglement cannot be used
for superluminal communication, has been phrased as ``peaceful
coexistence'' \cite {shimony} between quantum mechanics and
relativity, a much-cited expression. Here we emphasize that this
consistency is not a coincidence, but a simple consequence of
linearity and completeness of quantum mechanics. Our arguments go
beyond previous work \cite{grw,Herbert,Wootters,ghi,BuschA,svetlichny,Peres2},
as we consider the most general evolution on Alice's and Bob's side
in the form of local maps.

Recently, this consistency has been exploited in order to devise new
methods to derive bounds or constraints for quantum mechanical
transformations \cite{Gisin,Hardy,ghosh,Pati}.  However, in this paper
we will show that the  principles underlying the impossibility of
1) superluminal signalling and 2) quantum cloning beyond the optimal
bound allowed by quantum mechanics \cite{gima,oxibm,werner,bc,guo}, are
not the same.  In particular, the impossibility of information
transfer by means of quantum entanglement is due only to  linearity
and preservation of trace of local operations.

\section{Impossibility of superluminal communication}

In this section we want to show how the impossibility of superluminal
communication arises by assuming only completeness and
linearity of local maps on density operators.
  
We consider the most general scenario 
where Alice and Bob share a global quantum state $\rho_{AB}$ of two
particles and are allowed to perform any local map, which we
denote here with $\cala\otimes\eins$ and $\eins\otimes \calb$, 
respectively. The local map can be any local transformation, including
a measurement averaged over all possible outcomes (which, in fact,
cannot be known by the communication partner). 
Alice can choose among different local maps in order
to encode
 the message ``$m$'' that she wishes to transmit, namely
she encodes it by 
 performing the transformation
$\cala_m\otimes\eins$ on her particle. 
 Bob can perform a local
transformation $\eins\otimes\calb$ on his particle (e.g. cloning) 
 and then a local
measurement $\eins\otimes\Pi_r$ to decode the message 
 ($\Pi_r$ is a
POVM \cite{Helstrom,Peres}). The impossibility of superluminal
communication in the particular case where Bob performs only a
measurement has been demonstrated in Ref. \cite{grw}. Here we follow a
more general approach, discussing the roles of
``completeness'' and linearity of any local map involved. By
``completeness'' of a map \cala\ we mean that the trace is preserved under
its 
 action, namely
\begin{equation}
{\mbox{Tr}}[\cala(\rho_A)]\equiv {\mbox{Tr}}[\rho_A]
\end{equation}
for any $\rho$ \cite{nota}. Linearity of the map on trace-class operators 
of the form $\ket{\psi}\bra{\phi}$, allows to extend
the completeness condition to the whole Hilbert space, namely
\begin{equation}
{\mbox{Tr}}[\cala\otimes \eins(\rho_{AB})]\equiv {\mbox{Tr}}[\rho_{AB}]\;,
\end{equation}
and analogously for the partial trace
\begin{equation}
{\mbox{Tr}}_A[\cala\otimes \eins(\rho_{AB})]\equiv {\mbox{Tr}}_A[\rho_{AB}]
\label{part}\;,
\end{equation}
On Bob's side, only linearity without completeness is needed for the
local map \calb , leading to the equality
\begin{equation}
{\mbox{Tr}}_A[\cala\otimes \calb(\rho_{AB})]= 
\calb\,{\mbox{Tr}}_A[\cala\otimes\eins(\rho_{AB})]\;.\label{gcomp}
\end{equation}
As we will show in the following, the above equations are
the fundamental ingredients and the only requirements for local maps 
to prove the impossibility of superluminal communication. 
\par We will now compute the conditional probability $p(r|m)$
that Bob records the result $r$ when the message $m$ was encoded by Alice:
\begin{equation}
p(r|m)={\mbox{Tr}}[\eins\otimes \Pi_r(\cala_m\otimes \calb \;(\rho_{AB})) ]\;.
\end{equation}
By exploiting Eqs. (\ref{gcomp}) and (\ref{part}) we have
\begin{eqnarray}
p(r|m)&=&{\mbox{Tr}}_B[\Pi_r\,\calb\,
({\mbox{Tr}}_A[\cala _m\otimes\eins(\rho_{AB})])] 
\nonumber \\&= & 
{\mbox{Tr}}_B[\Pi_r\,\calb\,
({\mbox{Tr}}_A[\rho_{AB}])]\equiv p(r) 
\;.\label{gcomp2}
\end{eqnarray}
The conditional probability is therefore independent of the local
operation
 $\cala_m$ that Alice performed on her particle, and
therefore the amount of transmitted information vanishes.
Note that the speed of transmission does not enter in any way, i.e. 
{\em any} transmission of information is forbidden \cite{nota_asher}, 
in particular superluminal transmission. 

We want to stress that this result holds for all possible linear
local operations that Alice and Bob can perform, and also for any
joint state $\rho_{AB}$. In particular, it holds for any kind of
linear cloning  transformation performed at Bob's side (notice that
ideal cloning is a non-linear map). Notice also that any operation that
is physically realizable in standard quantum mechanics (completely
positive map) is  linear and complete, and therefore it does not
allow superluminal communication.  
\par
\begin{figure}[hbt]
\vskip .5truecm
\begin{center}\epsfxsize=.8 \hsize\leavevmode\epsffile{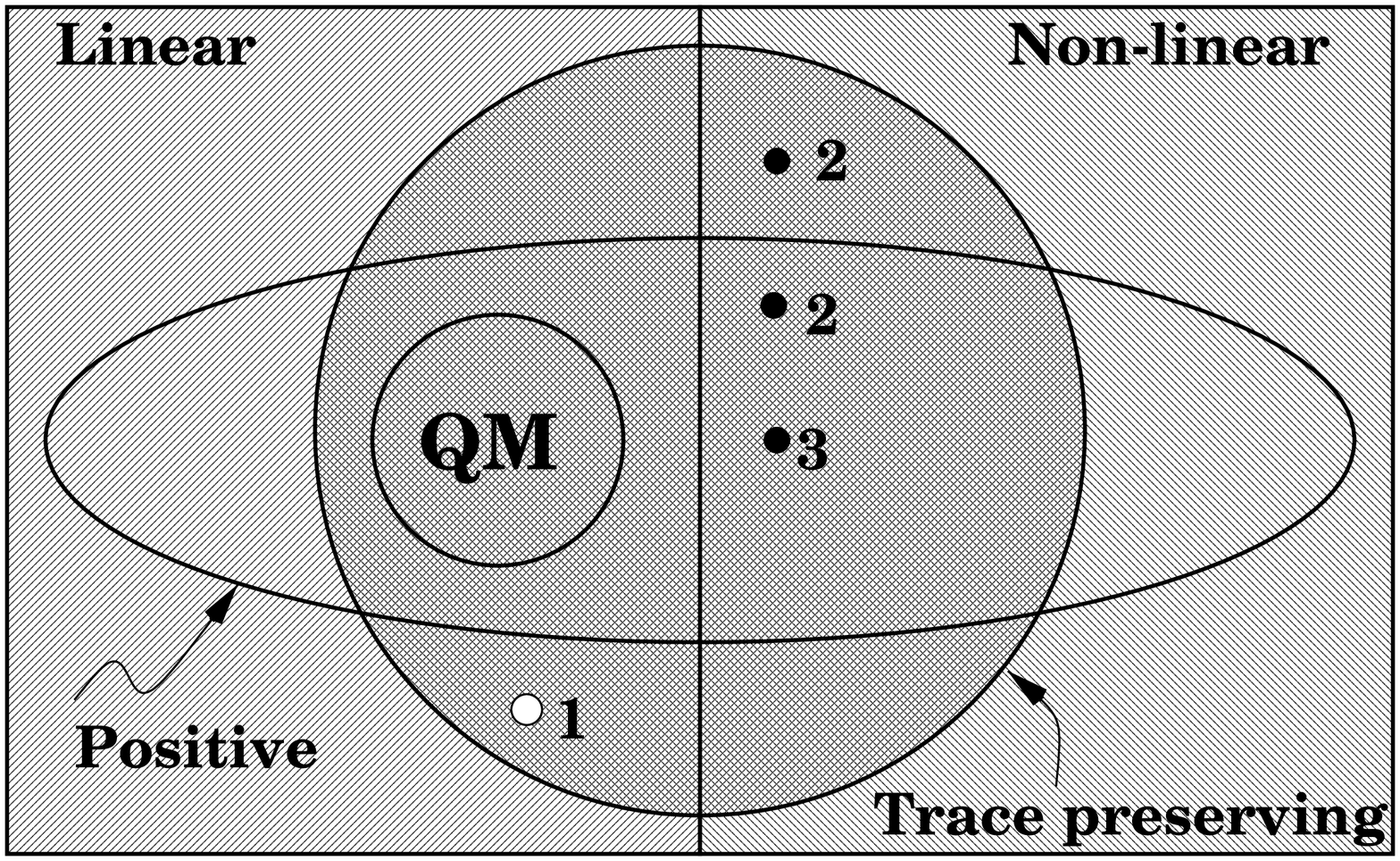}
\end{center}
\vskip .5truecm
\caption[]  
        {\small Diagram of local maps. QM denotes quantum mechanical maps, 
         namely linear trace-preserving CP maps \cite{nota}. Examples 
         from the text are placed in the diagram: map 1 (open circle) 
         does not allow superluminal communication; map 2 and 3 
         (full circle) do.}
\label{maps}
\end{figure}

 We  also emphasize here that
the ``peaceful coexistence'' between quantum mechanics and relativity
is automatically guaranteed by the linearity and 
 completeness of
any quantum mechanical process.
 Actually, as shown in the diagram
\ref{maps}, the set of local quantum mechanical
 maps is just a
subset of the local maps that do not allow superluminal 
communication.
In the next section we will show how
superluminal communication could be achieved if one would give up the
linearity requirement for the local maps, by discussing some explicit
examples.

\section{Examples}
\label{examples}

Our examples are based on the scenario where Alice and Bob share an entangled
state of two qubits and Alice performs a projection measurement with her 
basis oriented
along the direction $\vec{n}$. The final state of Bob, who does
not know the
 result of the measurement, is given by
 
\begin{equation}
p(\vec n)\rho_{out}(\vec n)+p(-\vec n)\rho_{out}(-\vec n)\;,\label{final}
\end{equation}
where $p(\pm\vec n)$ denote the probabilities that Alice finds her
qubit oriented as $\pm\vec n$, and $\rho_{out}(\pm\vec n)$ are the
corresponding final density operators at Bob's side after 
he performed his local
transformation.  Notice that the evolved state of Bob, as in the
following examples, can be a joint state of a composite system with
more than one qubit.  If the information is encoded in the choice of
two possible different orientations $\vec n_1$ and $\vec n_2$ of the
measurement basis, the impossibility of superluminal communication
corresponds to the condition
\begin{eqnarray}
&&p(\vec n_1)\rho_{out}(\vec n_1)+p(-\vec n_1)\rho_{out}(-\vec n_1)
\nonumber \\&&= 
p(\vec n_2)\rho_{out}(\vec n_2)+p(-\vec n_2)\rho_{out}(-\vec n_2)
\label{nsc}
\end{eqnarray}
for all choices of $\vec n_{1}$ and $\vec n_{2}$. 
In the following section we give some 
explicit examples of local maps on Bob's side.
Notice that we will
 intentionally leave the ground of quantum mechanics
(an explicit example of a superluminal communication scheme based on the
use of non-linear evolutions is also given in Ref. \cite{Gisin2}).
  
(1) {\em Example of a linear, non-positive $1\to 2$ cloning transformation
which
 does not allow superluminal communication:} 
\newline The
evolved state at Bob's side
 after his transformation is a state of
two qubits given by
\begin{eqnarray}
&&\rho_{out}(\vec s)= 
\nonumber \\&&\frac{1}{4}[\eins\otimes \eins
+\eta(\vec s\cdot\vec\sigma\otimes\eins+\eins\otimes\vec s\cdot\vec\sigma)
+t\sum_{j=x,y,z}\sigma_j\otimes\sigma_j]
\label{rhoout1}
\end{eqnarray}
where $\vec s$ 
is the Bloch vector which is cloned and $\eta$ is
the shrinking factor.
The above map is non-positive for $\eta>(1+t)/2$ \cite{Gisin}. 
This is the case,
for instance, for $t=1/3$
and $\eta>2/3$. Such a transformation violates  the upper 
bound of the $1\to 2$ universal 
quantum cloner \cite{bh,oxibm} -- but,
as this is a linear transformation, Eq. (\ref{gcomp2}) holds.
Therefore the cloning is ``better'' than the optimal one, and
the no-signalling condition (\ref{nsc})
is still fulfilled. 
\par
This means that we can go beyond the laws of quantum mechanics (complete
positivity)
without necessarily creating the possibility of superluminal communication.

(2) {\em Example of non-linear, positive or non-positive $1\to 2$ cloning
transformation which does allow superluminal communication:} \newline 
Consider Bob's transformation
\begin{eqnarray}
\rho_{out}(\vec s)&=&\frac{1}{4}[\eins\otimes \eins
\nonumber \\& + & 
(\sum_{j=x,y,z}f_j(s_j)\sigma_j\otimes\eins+
\eins\otimes\sum_{j=x,y,z}f_j(s_j)\sigma_j) \nonumber \\&  
+ & t\sum_{j=x,y,z}\sigma_j\otimes\sigma_j]\ ,
\label{rhoout2}
\end{eqnarray}
where $f_j(s_j)$ denotes a function of the component $j$ of the Bloch vector,
which is
such that this map acts non-linearly on a convex combination of density
matrices.
For odd functions, namely $f_j(s_j)=-f_j(-s_j)$ 
one does  not violate the no-signalling condition for
a maximally entangled state because taking $\vec s=\pm \vec n$ it
follows that 
$\rho_{out}(\vec n)+\rho_{out}(-\vec n)$ does not depend on $\vec n$,
whereas for even non-constant functions one does. However, for odd functions
the no-signalling condition is in general violated for partially entangled 
pure states, i.e. $p(\vec n)\neq p(-\vec n)$ in Eq. (\ref{final}). 
It is interesting to see that in this non-physical case 
superluminal communication is achieved when sharing less than
maximal entanglement.
\par
Depending on the value of the parameter $t$ this map can be positive or 
non-positive.
Examples of non-positive maps can for instance
be found by violating the condition
$f_z(1)>(1+t)/2$ (compare with previous example).

(3) {\em Example of a 
non-linear, positive  $1\to N$ cloning transformation which
does allow superluminal communication:} \newline
Consider
\begin{eqnarray}
&&\proj{\psi}\otimes \proj{0}^{\otimes (N-1)}
   \to \nonumber \\& & 
F \proj{\psi}^{\otimes N}+(1-F)\proj{\psi_\perp}^{\otimes N}\;,
   \ \ \ N\geq 2 \ ,
   \label{nonlinear}
\end{eqnarray}
where $\ket{\psi_\perp}$ is orthogonal to $\ket{\psi}$.
The no-signalling condition (\ref{nsc})
for two different choices of basis $\{\ket{\psi},
\ket{\psi_\perp}\}$ and $\{\ket{\phi},\ket{\phi_\perp}\}$ with
equiprobable outcomes  is violated because
\begin{equation}
\proj{\psi}^{\otimes N}+\proj{\psi_\perp}^{\otimes N}\neq
\proj{\phi}^{\otimes N}+\proj{\phi_\perp}^{\otimes N}\;,
\label{dist}
\end{equation}
which holds for any value $0\leq F\leq 1$.
 It is then possible to devise a
measurement procedure that distinguishes
 between the left and right
hand side of Eq. (\ref{dist}), thus allowing to transmit 
 information faster than
light.  
\par In order to illustrate this we give an explicit example with
$N=2$.
 Let us denote the right hand side of equation
(\ref{nonlinear}) as $\bar\rho(\psi)$.  We choose $\ket{\psi}=\ket{0}$
and $\ket{\phi}=(\ket{0}+\ket{1})/\sqrt{2}$
 and a POVM measurement
on the clones given by the operators $E_0$ and $E_1$, which are
 the
projectors over the subspaces spanned by $\{\ket{01},\ket{10}\}$ and
$\{\ket{00},\ket{11}\}$, respectively.
\par With this measurement the
probabilities for outcome 0 and 1 depend on Alice's choice of
measurement basis. We denote as $p(0|\psi)$ the probability that Bob
finds outcome 0, if Alice measured in the basis
$\{\ket{\psi},\ket{\psi_\perp}\}$, and arrive at
\begin{eqnarray}
p(0|\psi)&=&\frac{1}{2}{\mbox {Tr}} [E_0(\proj{\psi}^{\otimes 2}
       +\proj{\psi_\perp})^{\otimes 2}]=0 \;,\nonumber \\
p(1|\psi)&=&1-p(0|\psi)=1 \;.
\end{eqnarray}
Analogously, for the other choice of Alice's basis one has
\begin{eqnarray}
p(0|\phi)&=&
\frac{1}{2}{\mbox {Tr}} [E_0(\proj{\phi}^{\otimes 2}
      +\proj{\phi_\perp}^{\otimes 2})]=\frac{1}{2} \;,\nonumber \\
 p(1|\phi)&=&1-p(0|\phi)=\frac{1}{2}     
      \;.
\end{eqnarray}
Therefore, we can distinguish between the two different choices of
bases. Note that, when giving up the constraint of linearity, 
one could send signals superluminally even for fidelities smaller
than those of optimal quantum cloning.
\par\noindent Similar arguments hold  for the transformation
\begin{eqnarray}
&&\proj{\psi}\otimes \proj{0}^{\otimes (N-1)}
  \to \nonumber \\& &    \left(F \proj{\psi}+(1-F)\proj{\psi_\perp}\right)
^{\otimes N}\;.
\label{factor}
\end{eqnarray}

\section{Conclusions}
We have shown that the ``peaceful coexistence'' between quantum
mechanics  and relativity is automatically guaranteed by the
linearity and completeness (i.e. trace-preserving property) of any
quantum mechanical process: hence, any approximate optimal quantum
cloning, as a particular case of a linear trace-preserving map, cannot
lead to signalling. 
\par For the sake of illustration, in figure \ref{maps} we summarize
the set of local maps.  
This set is divided into linear and non-linear maps. Any linear
trace-preserving map forbids superluminal signalling. Reversely, the
no-signalling condition implies only linearity, as shown in
Refs.\cite{svetlichny} and \cite{Gisin2,Gisin3}. The positive maps
contain the linear maps allowed by quantum mechanics (QM), namely the
completely positive trace-preserving maps. Both trace-preservation and 
positivity---crucial for quantum mechanics---are not implied by
the no-signalling constraint. In particular, positivity seems to 
be unrelated with no-signalling. 
Hence, there is room for maps that go beyond quantum mechanics, but
still preserve the constraint of no-superluminal signalling, and Example 1)
above shows that this is the case.

\par From what we have seen we can conclude that any bound on a
cloning fidelity cannot be derived from the no-signalling constraint
alone, but only in connection with other quantum mechanical
principles: Example 3) shows how the cloning fidelity is unrelated to
the no-signalling condition. Quantum mechanics as a complete theory,
however, naturally guarantees no-signalling, and obviously gives the 
correct known upper bounds on quantum cloning.

\section{Acknowledgements}
We thank C. Fuchs, G. C. Ghirardi, L. Hardy and A. Peres for fruitful
discussions. 
DB acknowledges support by  the ESF Programme QIT,
and from Deu\-tsche For\-schungs\-ge\-mein\-schaft  under SFB 407 and 
Schwer\-punkt QIV. 
The Theoretical Quantum Optics Group of Pavia
acknowledges the European Network EQUIP and Cofinanziamento 1999 
``Quantum information transmission and processing: quantum
teleportation and error correction'' for partial support.


\begin{thebibliography}{99}
\bibitem{nota} We refer to the map \cala\ as complete as synonymous of 
trace-preserving since generally Alice's map is a measurement, and
summing over all possible outcomes (i.e. for the completeness of the
measurement) leads to a linear trace-preserving map (non linear state 
reduction maps are always linear on average).
\end{thebibliography}
\end{document}